\documentclass[letterpaper,12pt]{article}   %% LaTeX 2e (preferred)
\usepackage{osajnl2} %% do not use with REVTeX4
\usepackage{epsfig}
\begin{document}
\title{Self-consistent phase determination for Wigner function reconstruction}
\author{Maria Bondani,$^{1,*}$ Alessia Allevi,$^2$ and Alessandra Andreoni$^{3}$}
\address{$^1$National Laboratory for Ultrafast and Ultraintense
Optical Science - C.N.R.-I.N.F.M. and C.N.I.S.M., U.d.R. Como,  I-22100, Como, Italy}
\address{$^2$C.N.I.S.M., U.d.R.  Milano, I-20133, Milano, Italy}
\address{$^3$Dipartimento di Fisica e
Matematica, Universit\`a degli Studi dell'Insubria and C.N.I.S.M., U.d.R. Como, \\ I-22100, Como, Italy}
\address{$^*$Corresponding author: maria.bondani@uninsubria.it}

%%%%%%%%%%%%%%
\begin{abstract}
We present the reconstruction of the Wigner function of a classical phase-sensitive state, a pulsed coherent state, by measurements of the distributions of detected-photons of the state displaced by a coherent probe field. By using a hybrid photodetector operated above its photon-resolving regime, we obtained both the statistics at different values of the probe field and the values of the probe phase required to reconstruct the Wigner function.
\end{abstract}
%%%%%%%%%%%%%%%%%%%%%%%%%%%%%%%%%%%%%%%%%%%%%%%%%%%%%%%
% ] %% activate for two-column option
\maketitle
%%%%%%%%%%%%%%%%%%%%%%%%%%%%%%%%%%%%%%%%%%%%%%%%%%%%%%%%%%%%%
\section{Introduction}\label{sec:intro}
Quantum optical states are a valuable resource in view of the development of quantum protocols as the many degrees of freedom of the radiation field can be exploited to enhance transmission, manipulation and storage of quantum information.
To this aim, during the last decades many efforts have been devoted to the generation of suitable quantum states of light: besides the production of different kinds of squeezed states \cite{Wenger04A,Vahlbrouch}, conditional states endowed with non-Gaussian nature have been produced starting from nonlinear parametric processes \cite{Wenger04B,Ourjoumtsev06,Ourjoumtsev07,Zavatta07,Bondani07}. Among them phase-sensitive quantum states can be considered a good resource as they also carry phase information \cite{Lvovsky02}. Obviously, to exploit such states reliable characterization strategies are required.
The typical technique employed so far is quantum homodyne tomography (OHT) \cite{Lvovsky09} operating either in spectral or in time domain \cite{Smithey93}. OHT was proposed in 1989 by Vogel and Risken \cite{Vogel} and is based on balanced homodyne detection \cite{Yuen83,Abbas83}, where the interference of a signal field and a local oscillator (LO), a coherent field with variable phase, provides the values of the quadrature amplitude distributions of the signal as a function of the LO phase. In general the variation of the phase is obtained by means of piezoelectric transducers controlled by a feedback system. In order to overcome the unwanted fluctuations arising from the instabilities of the movable elements, quite recently high-frequency time-domain OHT has been proposed and implemented \cite{Lvovskyhomo,Wenger05,ZavattaLPL}.
This technique can operate at the high repetition rate available from mode-locked lasers. Unfortunately, to obtain quantum states containing sizeable numbers of photons, low rep-rate laser sources must be used, and the acquisition of OHT data may become unpractically long and affected by uncontrollable instabilities.\\
To overcome this problem, here we propose a self-consistent Wigner function reconstruction based on the interferometric scheme depicted in Fig.~\ref{f:setup} \cite{Cahill}. The method consists in detecting the light exiting a beam-splitter that mixes the signal field to be characterized with a coherent probe field, whose amplitude, $|\alpha|$, and phase, $\phi$, can be varied with continuity.
We have already applied this method to the measurement of the Wigner function of some phase-insensitive fields \cite{OL} produced by a low rep-rate source (15~kHz); now we present a measurement on a phase-sensitive coherent field from the same source. At such a rate, the measurements took a long time and the problem of determining the phase in the presence of potential instabilities was solved by exploiting the linearity of the detector in a completely self-consistent analysis procedure.

The paper is organized as follows: in Section~\ref{sec:teo} we present the theoretical background of the method, in Section~\ref{sec:exp} we present in detail experimental procedure and results and finally we conclude in Section~\ref{sec:concl}.

%%%%%%%%%%%%%%%%%%%%%%%%%%%%%%%%%%%%%%%%%%%
\section{Theory}\label{sec:teo}

The Wigner function of a single mode state can be written as \cite{Cahill}
\begin{eqnarray}
 W(\alpha) = \frac{2}{\pi} \sum_{n=0}^\infty (-1)^n p_{n}(\alpha)\; ,
\label{eq:wigner}
\end{eqnarray}
where $p_{n}(\alpha)$ is the photon-number distribution of the field displaced by $\alpha$, which is not directly accessible by using real detectors having quantum efficiency $\eta<1$.\\
%%%%%%%%%%
However, if we are able to characterize the detector, knowing the statistics of detected photons can be enough for a full characterization of the signal field. For instance, if we assume that the primary detection process is described by a Bernoullian convolution, we have that the detected-photon distribution, $\overline{p}_{m}(\beta)$, is given by \cite{mandel1995}
\begin{equation}
 \overline{p}_{m}(\beta)=\sum_{n=m}^{\infty} \left(
 \begin{array}{c}n\\m\end{array}\right)
 \eta^m (1-\eta)^{n-m} p_{n}(\alpha)\ ,\label{eq:phel}
\end{equation}
where $\beta=\sqrt{\eta}\alpha$ is the detected amplitude of the displacement field.\\
In principle, $\overline{p}_{m}(\beta)\neq  p_{n}(\alpha)$, that is the distribution of the photoelectrons is not necessarily the same as that of the photons. Nevertheless it can be demonstrated \cite{ASL,casini} that when the state can be obtained by combining non-squeezed gaussian fields, the statistics of detected photons remains the same. \\
If Eq.~(\ref{eq:phel}) holds, we can write for the Wigner function of the detected photons an expression analogous to that in Eq.~(\ref{eq:wigner}) \cite{OL}
\begin{equation}
 \overline{W}(\beta) =\frac{2}{\pi} \sum_{m=0}^\infty (-1)^m \overline{p}_{m}(\beta)\ ,\label{eq:wigEL}
\end{equation}
where
\begin{eqnarray}
 \overline{W}(\beta) =\frac{2}{\pi (1-\eta)}\int d^2\beta' e^{-\frac{2}{1-\eta}|\beta-\beta'|^2} W(\beta'/\sqrt{\eta})\; .
\label{eq:wigLOSS}
\end{eqnarray}
is the Wigner function in the presence of losses \cite{Banas96} and $W(\beta/\sqrt{\eta})=W(\alpha)$ is that of the photons. Obviously, if $\overline{p}_{m}(\beta)$ has the same functional form as $p_{n}(\alpha)$, also the Wigner function remains the same.\\
As an axample, in the case of a coherent signal, the Wigner function for the detected state is
\begin{eqnarray}
 \overline{W}(\beta) = \frac{2}{\pi} \exp\left(-2|\beta - \beta_0|^2\right)\; ,
\label{eq:wignergauss}
\end{eqnarray}
where $\beta_0$ is the complex amplitude of the considered coherent signal field.

The statistics of detected photons can be obtained with some classes of detectors, including photomultipliers \cite{Arecchi,Burle}, cryogenic photon counters \cite{Yamamoto99,Waks04}, multi-pixel photon counters \cite{Hamamatsu}, time-multiplexed detectors \cite{Achilles,Laiho}, multichannel fiber loop detectors \cite{Rehacek}. Obtaining or not the statistics of detected photons depend on the photon-counting capability of the detector, a fact that limits the range of intensities that can be investigated.
For all these reasons, the only experimental Wigner functions directly measured by photon counting are those reported in Ref.~\cite{Banas99} for very low intensities.\\
Recently we introduced a new method to analyze the output signal of linear detectors, that allowed us to reconstruct the detected photon distribution for a number of optical states without any \emph{a priori} knowledge  \cite{JMO,ASL,andreoniPRA}. In our technique we consider a linear photodetector endowed with high gain and model the detection process through a Bernoullian convolution \cite{Mandel1986} and the overall amplification and conversion process through a very precise factor, which is taken as constant, $\gamma$ \cite{andreoniPRA}. With these assumptions, we can link the statistics of the detector output (voltages in the present case) to that of the detected photons. In particular, we can write the mean value as $\overline{v} = \gamma \overline{m}=\eta\gamma\overline{n}$, $\eta$ being the overall detection efficiency, and the variance as $\sigma^2(v) = \gamma^2\sigma^2(m)=\gamma^2(\sigma^2(n)+ \eta(1-\eta)\overline{n})$. We can thus measure the Fano factor of the output voltages, $F_v=\sigma^2(v)/\overline{v}$, at different values of $\eta$. We observe that $F_v$ can be written in the very general form
\begin{eqnarray}
 F_v  = \frac{Q}{\overline{n}} {\overline{v}} + \gamma\; ,
\label{eq:fanoQ}
\end{eqnarray}
where all the dependence on the field under investigation is in the angular coefficient, $Q / \bar{n}$, $Q$ being the Mandel parameter. We can thus obtain $\gamma$ from a fit of experimental data. Once $\gamma$ is evaluated, it is possible to find the photoelectron distribution by dividing the $v$ output values by the experimental value of $\gamma$ and re-binning the data in unitary bins.
Our method has the advantage of being self-consistent as the value of $\gamma$ is obtained from measurements on the same field under investigation \cite{JMO} and does not require independent calibration.
%%%%%%%%%%

%%%%%%%%%%
\section{Experiment}\label{sec:exp}

\begin{figure}[h]\label{f:setup}
  \centering
  \includegraphics[angle=0,width=0.5\textwidth]{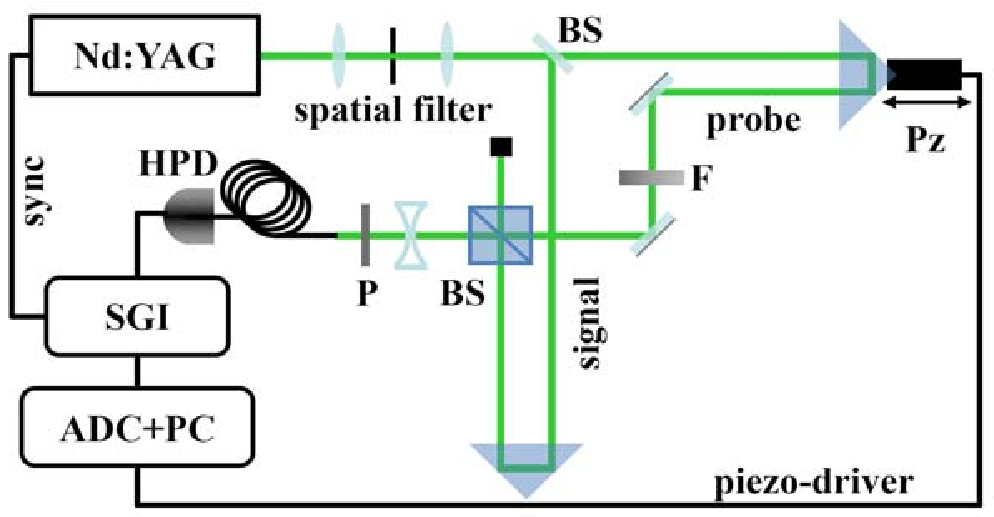}
  \caption{(Color online) Experimental setup. HPD, hybrid photo-detector; BS, beam splitter; F, neutral density filter; Pz, piezoelectric movement; P, polarizer; SGI, synchronous gated integrator.}
  \end{figure}

Aim of the experiment is to reconstruct the Wigner function of a phase-sensitive state (signal), a coherent state, by using Eq.~(\ref{eq:wignergauss}). To reach this goal, the probability distribution of detected photons must be measured at different values of the complex displacement, $\alpha = |\alpha|\exp(i\phi)$ (probe), to obtain $\overline{p}_m(|\beta|\exp(i\phi)=\sqrt{\eta}|\alpha|\exp(i\phi))$.
The experimental setup is sketched in Fig.~\ref{f:setup}.\\
The light source was a frequency-doubled Q-switched Nd:YAG laser operating at 15~kHz repetition rate (Quanta System) and providing linearly polarized pulses at 532~nm of $\sim200$~ns duration. The beam was spatially filtered and split into two parts serving as signal and probe fields. Intensity and phase of the probe field were adjusted by means of a variable neutral filter (F) and of a piezoelectric movement (Pz), respectively. Signal and probe were then mixed at a cube beam-splitter (BS in Fig.~\ref{f:setup}). The field exiting one of the ports of the BS was passed through a diverging lens and a portion of it was collected by a multimode optical fiber (600~$\mu$m core-diameter) and delivered to a hybrid photodiode module (HPD, H8236-40 with maximum quantum efficiency $\sim0.4$ at 550~nm, Hamamatsu) operated above its photon-resolving regime. In fact, this detector is endowed with a limited photon-number resolving capability but is linear over a wide range of intensities.\\
We sampled the phase space by changing  amplitude and phase of the probe independently. In particular, we considered  46 values of probe intensity, $|\beta|^2$, set by adjusting the variable filter F, and 48 values of phase, $\phi$, at each $|\beta|$, set by moving the Pz in steps.
Before starting the piezoelectric movement, we made independent measurements of both signal and probe, and evaluated their mean values $|\beta_0|^2$ and $|\beta|^2$, respectively.
At each one of the 46$\times$48-values of the probe, we performed the calibration procedure described in Section~\ref{sec:teo}, to obtain the value of $\gamma$. Each calibration run included 25 $\eta$-values, set by means of the polarizer P in Fig.~\ref{f:setup} from the maximum value $\eta_{max}=0.31$ down to almost zero. For each $\eta$ we measured the output voltages, $v$, at 30000 single-shots, we evaluated the Fano factor, $F_v$, and from its fit as a function of $\overline{v}$ we obtained the value of $\gamma$ according to Eq.~(\ref{eq:fanoQ}).
  \begin{figure}[h]\label{f:proced}
  \centering
  \includegraphics[width=0.5\textwidth]{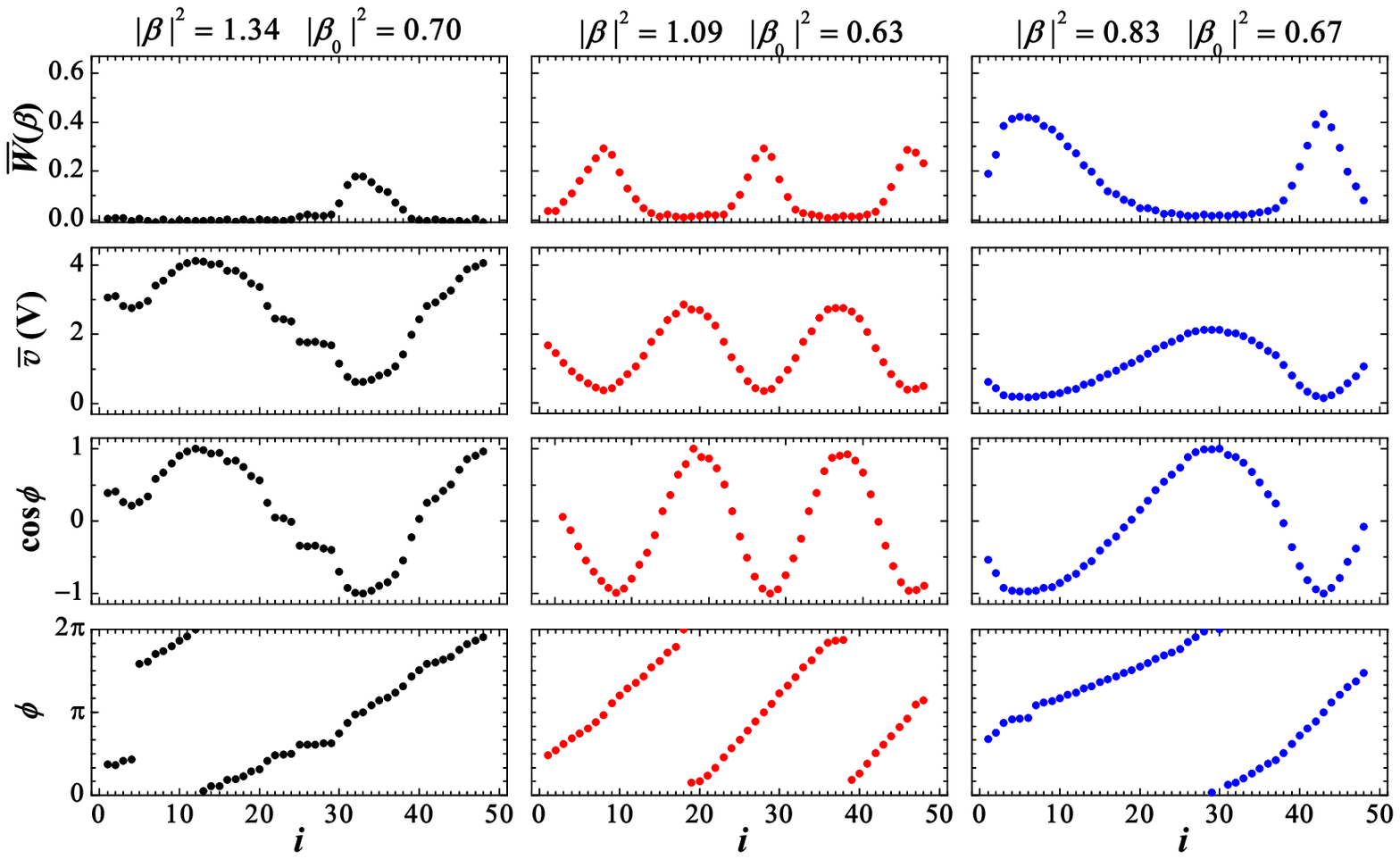}
  \caption{(Color online) Columns: measurements taken at the indicated values of probe, $|\beta|^2$, and signal, $|\beta_0|^2$, intensities. First row: Wigner function reconstructed according to Eq.~(\ref{eq:wigEL}) as a function of the piezo step, $i$. Second row: mean value of the output voltages at $\eta_{max}$. Third row: experimental cosine of the interference pattern. Fourth row: phase values obtained from the cosine.}
  \end{figure}
By using this value of $\gamma$, we calculated the probability distributions $\overline{p}_m(\beta)$ corresponding to $\eta_{max}$ in each run and used them to calculate $\overline{W}(\beta)$ according to Eq.~(\ref{eq:wigEL}). The resulting values of the Wigner functions are plotted as a function of the step of the piezo, $i$, in the first row of Fig.~\ref{f:proced}. We observe that, due to unavoidable instabilities, the reproducibility of the movement of the piezo was not guaranteed. As a consequence, the Wigner functions thus reconstructed cannot be directly interpreted. Moreover, although a large part of the experimental procedure was automated, the complete acquisition required very long measurement sessions, performed in different days. For all these reasons, we had to devise a strategy to recover the phase values directly from the measurement by taking advantage of the linearity of the detector.\\
The procedure is outlined in the following. At each step of the piezo movement, we calculated the mean value, $\overline{v}$, of the most intense acquisition, corresponding to $\eta_{max}$ (see second row in Fig.~\ref{f:proced}). For regular steps of the piezo, we would expect a sinusoidal trend reproducing the interference pattern between signal and probe, but, in the absence of a feedback control on the piezo, its positioning was extremely irregular. Nevertheless, the pattern still represents an interference and thus we used the data to infer fringe visibility, and hence overlap of signal and probe, and phase difference between signal and probe.
We calculated the interference visibility as  $V = (\max(\overline{v})-\min(\overline{v}))/ (\max(\overline{v})+\min(\overline{v}))$ and the overlap as $\xi=V_{max}/(2-V_{max})$ \cite{Banas2002}, where $V_{max}$ was obtained when the intensity of the probe was equal to that of the signal.
In our system we had a limited visibility, $V_{max}=0.87$, and hence $\xi\sim0.78$. The visibility values for the data reported in Fig.~\ref{f:proced} were $V=0.73$,  $V=0.77$ and $V=0.86$. To obtain the phase, $\phi$, we evaluated a quantity that can be interpreted as a cosine of the phase, namely  $\cos(\phi) = (\overline{v}-(\max(\overline{v})+\min(\overline{v}))/2)/ (\max(\overline{v})+\min(\overline{v}))/2$ (see third row in Fig.~\ref{f:proced}). By inverting the cosine we obtained the values of $\phi$ displayed in the fourth row of the same figure. This method allows assigning the actual value of phase at each piezo position independent of the regularity and reproducibility of the movement.\\
In Fig.~\ref{f:wigPHASE} we present as dots the values of the experimental Wigner function, same data as in Fig.~\ref{f:proced}), plotted as a function of the obtained $\phi$ values.
 \begin{figure}[h]\label{f:wigPHASE}
 \centering
 \includegraphics[width=0.5\textwidth]{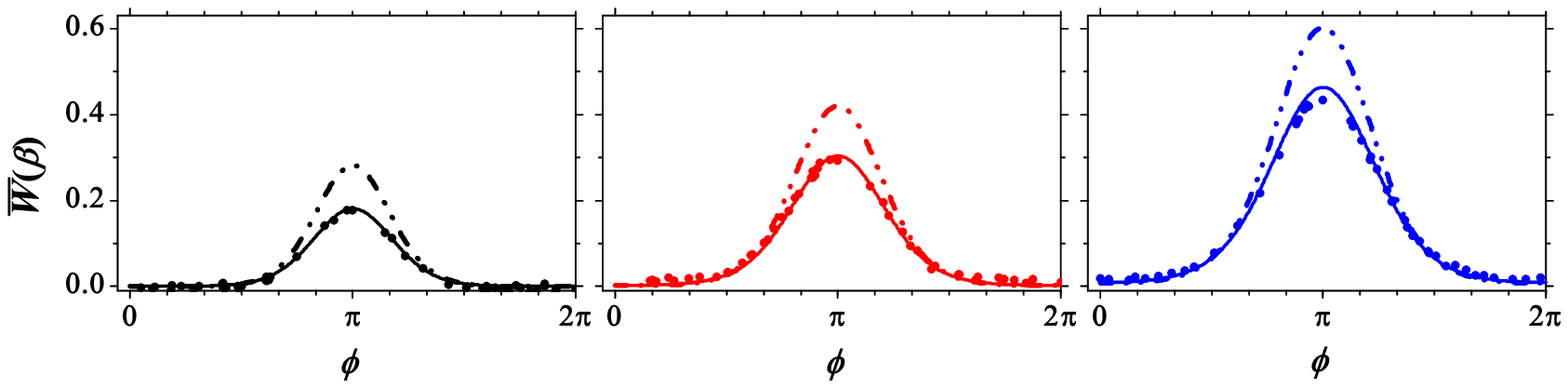}
 \caption{ (Color online) Dots: experimental reconstruction of sections of the Wigner function at fixed values of probe, $|\beta|^2$, and signal, $|\beta_0|^2$, intensities and variable $\phi$; dash-dotted lines: expected theoretical curves; full lines: theoretical curves corrected for the overlap of signal and probe.}
 \end{figure}
Together with the experimental data, in Fig.~\ref{f:wigPHASE} we also plot the expected theoretical curve evaluated from Eq.~(\ref{eq:wignergauss}) at the measured values $|\beta|$ and $|\beta_0|$ (dash-dotted line). The apparent discrepancy between theory and experiment can be ascribed to the non-perfect overlap, $\xi$, of signal and probe: only a part of the measured state comes from the superposition while the remaining is simply a spurious coherent contribution coming from a part of the probe. As a result, the measured distributions become the convolution of those expected and the poissonian distribution of the residual coherent probe \cite{OL}. In the present case, in which all the distributions are poissonian, no variation is visible in the $p_m(\beta)$, while the theoretical Wigner function must be recalculated as the product of the Wigner function expected for a signal reduced by the overlap with the gaussian Wigner function of the residual field \cite{Banas99}:
\begin{eqnarray}
 \overline{W}(\beta) = \frac{2}{\pi} \exp\left(-2(1-\xi)|\beta|^2\right) \exp\left(-2|\sqrt{\xi}\beta - \beta_0|^2\right)\; .
\label{eq:wignerREDUCED}
\end{eqnarray}
The theoretical Wigner functions, calculated from Eq.~(\ref{eq:wignerREDUCED}) with $\xi = 0.78$ and $|\beta_0| = 0.70$, $|\beta_0| = 0.63$ and $|\beta_0| = 0.67$, respectively, superimpose to the experimental data (see full lines in Fig.~\ref{f:wigPHASE}).\\
 \begin{figure}[h]\label{f:wig3D}
 \centering
 \includegraphics[width=0.5\textwidth]{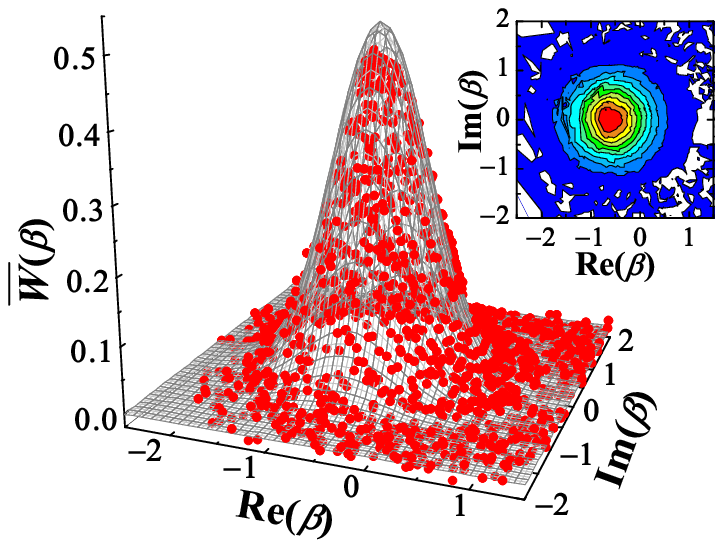}
 \end{figure}
\noindent
Fig. 4. (Color online) Main: experimental reconstruction of the Wigner function (dots) and theoretical prediction. Inset: contour plot of the experimental data.
Finally, Fig.~\ref{f:wig3D} shows the 3D reconstruction of the Wigner function obtained by combining all the measurements at different $\beta$ (dots).
The theoretical surface was obtained by correcting the expected Wigner function in Eq.~(\ref{eq:wignergauss}), taking into account the overlap and setting the mean value of the signal intensity at $|\beta_0|=0.67$.
For a quantitative estimation of the quality of the reconstruction, we evaluate the mean error $\epsilon = \sum_{k=1}^K [\overline{W}^{\rm th}(\beta_k)-\overline{W}^{\rm exp}(\beta_k)]/N$. The presented data yielded $\epsilon=-1.7\times10^{-3}$. As we can see, the theoretical surface carefully fits the experimental data in spite of all the experimental instabilities, thus testifying the robustness of both setup and analysis method.
%%%%%%%%%%%%%%%%%%%%%%%%%%%%%%%%%%%%%%%%%%%%%%%%%
\section{Conclusions}\label{sec:concl}
In conclusion, we implemented the reconstruction of the Wigner function of a pulsed, phase-sensitive, coherent field state containing a low, but non-negligible, number of photons ($\overline{n}=\overline{m}/\eta_{max}=2.12$). The technique takes advantage of the features of a hybrid photodetector, which is both able to distinguish a single detected photon and endowed with linear response over a wide intensity range. The heart of our experiment is the self-consistent procedure employed to analyze the detector output that allowed us to establish the statistics of detected fields and to assign the values to the phase of the probe field. The technique results to be reliable in spite of the low rep-rate of the laser source that imposed long experimental sessions to collect the data. This opens the way to the characterization of non-classical states, such as conditional states generated starting from twin-beam and squeezed states containing a sizeable number of photons, that can only be generated by low-rep-rate sources. Moreover, the self-consistent calibration procedure, which in the present work was repeated at each of the $46\times48$ values of the probe, as expected yielded almost constant values of $\gamma$ once we had set the parameters of the acquisition chain. Such a stability in $\gamma$ should allow us to perform the calibration procedure only at the beginning of the measurement, thus shortening the overall measurement time.

%%%%%%%%%%%%%%%%%%%%%%%%%%%%%%%%%%%%%%%%%%%%%%%%%

%%%%%%%%%%%%%%%%%%%%%%%%%%%%%%%%%%%%%%%%%%%%%%%%%%%%%%%%%%%%%%%%%%%%%%%%%%
\section*{Acknowledgments}
The Authors thank P. Salvadeo and A. Agliati (Quanta System S.p.A., Solbiate Olona, Italy) for the long-term loan of the laser and for the promptness of their technical assistance.\\
Present address of A. Allevi is: C.N.I.S.M., U.d.R. Milano, I-20133, Milano, Italy.
%%%%%%%%%%%%%%%%%%%%%%%%%%%%%%%%%%%%%%%%%%%%%%%%%%%%%%%%%%%%%%%%%%%%%%%%%

%%%%%%%%%

\clearpage

%%%%%%%%%%%%%%%%%%%%%%%%%%%%%%%%%%%%%%%%%%%%%%%%%
\section*{List of Figure Captions}

  \begin{figure}[h]
  \centering
  \includegraphics[angle=0,width=0.8\textwidth]{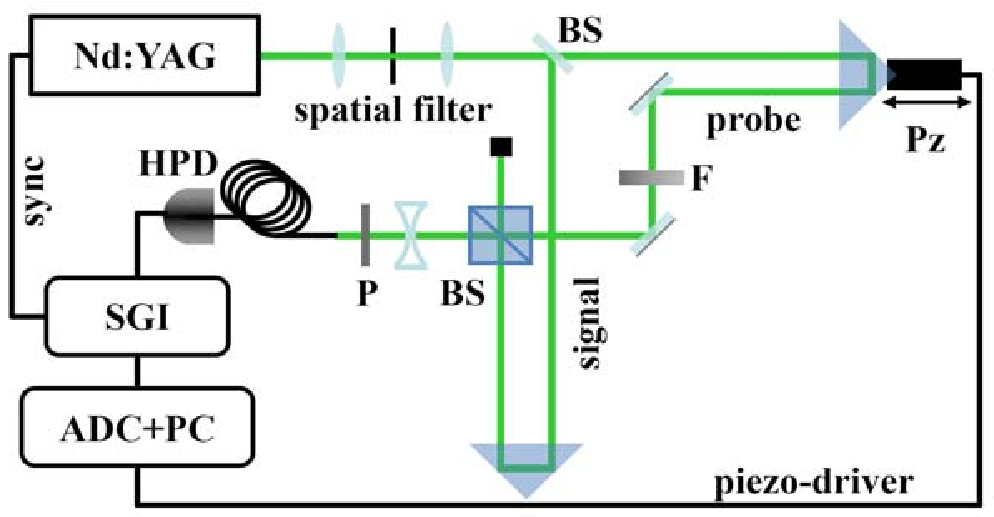}
  \end{figure}
\noindent
Fig. 1. (Color online) Experimental setup. HPD, hybrid photo-detector; BS, beam splitter; F, neutral density filter; Pz, piezoelectric movement; P, polarizer; SGI, synchronous gated integrator.
\clearpage
  \begin{figure}[h]
  \centering
  \includegraphics[width=0.8\textwidth]{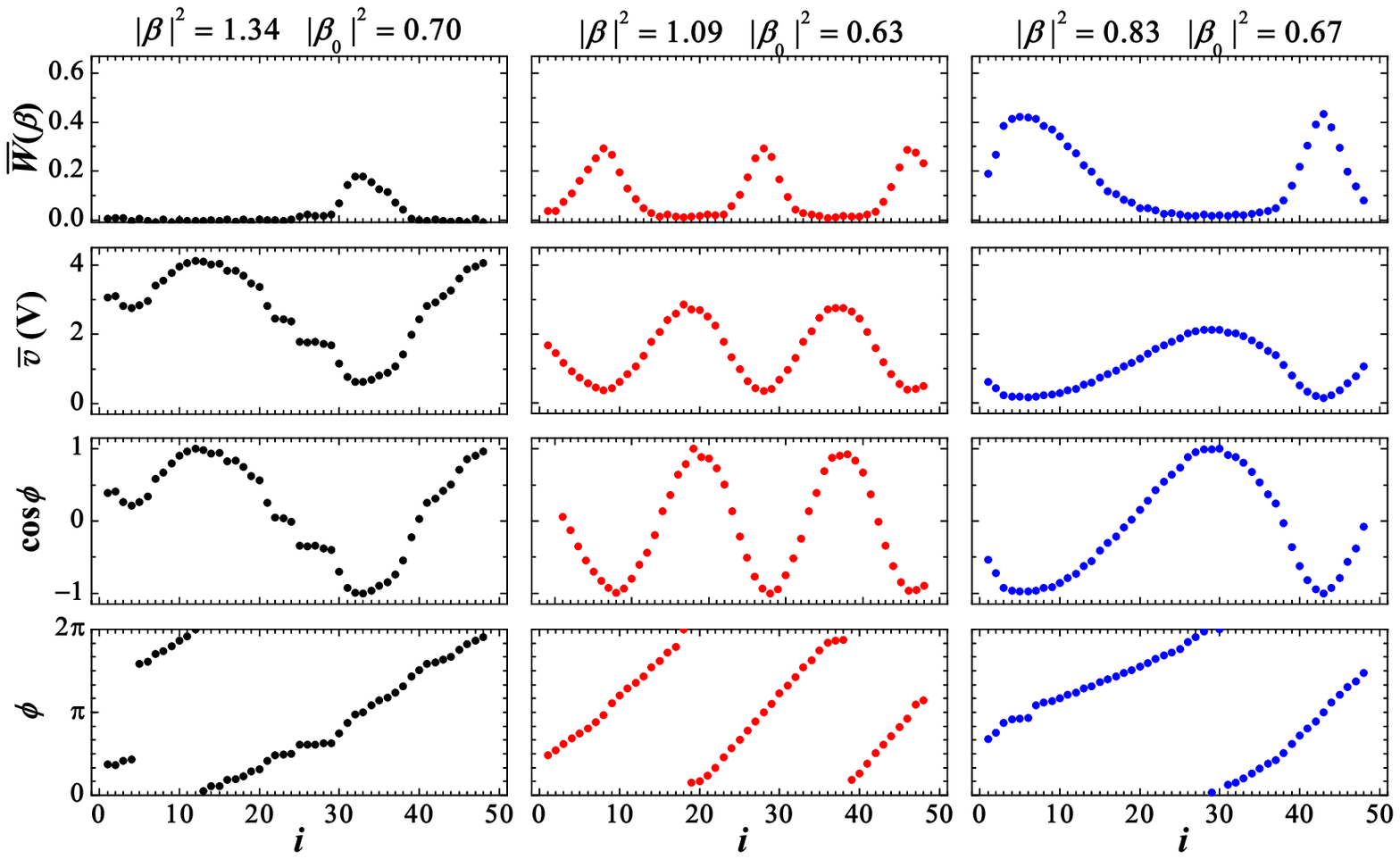}
  \end{figure}
\noindent
Fig. 2. (Color online) Columns: measurements taken at the indicated values of probe, $|\beta|^2$, and signal, $|\beta_0|^2$, intensities. First row: Wigner function reconstructed according to Eq.~(\ref{eq:wigEL}) as a function of the piezo step, $i$. Second row: mean value of the output voltages at $\eta_{max}$. Third row: experimental cosine of the interference pattern. Fourth row: phase values obtained from the cosine.
\clearpage
 \begin{figure}[h]
 \centering
 \includegraphics[width=0.8\textwidth]{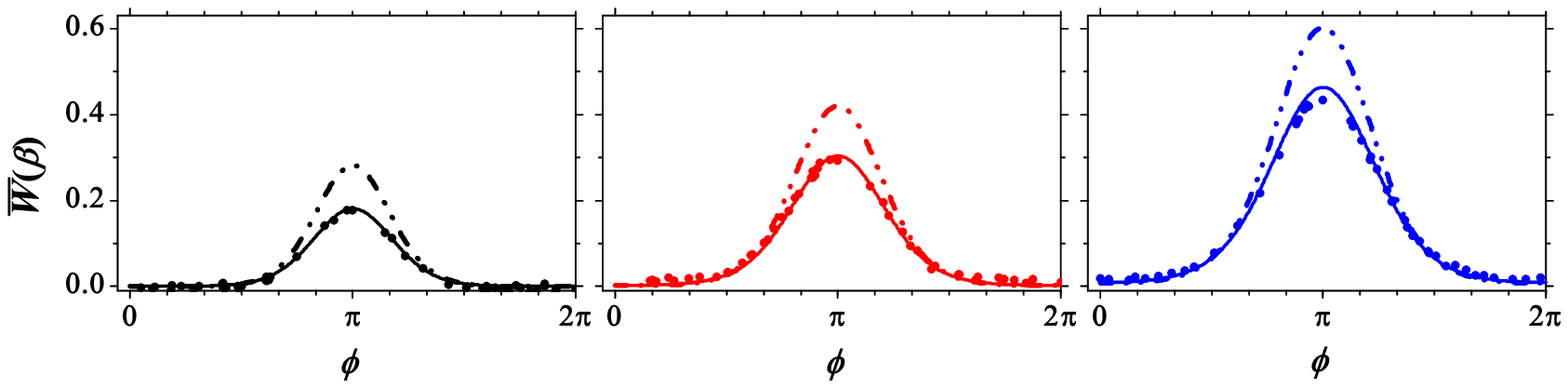}
 \end{figure}
\noindent
Fig. 3 (Color online) Dots: experimental reconstruction of sections of the Wigner function at fixed values of probe, $|\beta|^2$, and signal, $|\beta_0|^2$, intensities and variable $\phi$; dash-dotted lines: expected theoretical curves; full lines: theoretical curves corrected for the overlap of signal and probe.
\clearpage
\begin{figure}[h]
  \centering
  \includegraphics[width=0.8\textwidth]{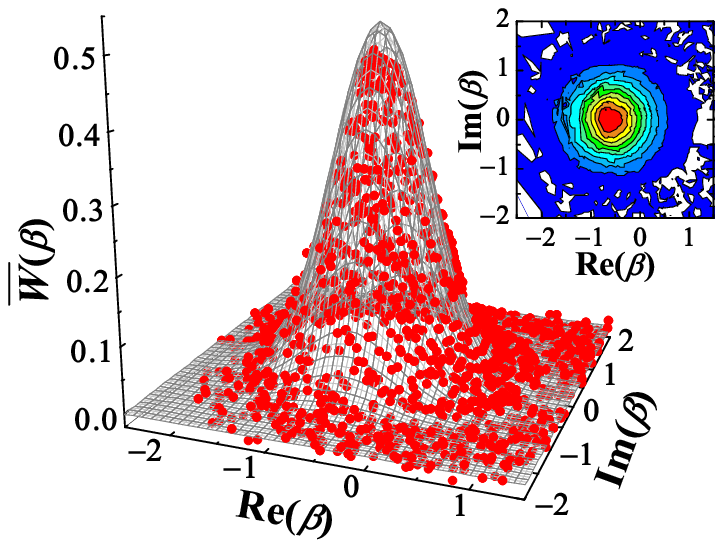}
  \end{figure}
\noindent
Fig. 4. (Color online) Main: experimental reconstruction of the Wigner function (dots) and theoretical prediction. Inset: contour plot of the experimental data.

\end{document}